\documentstyle[12pt]{article}
\input epsf
\input psfig
\newcommand{\nn}{\nonumber}
\newcommand{\g}{{\rm g}} 
\newcommand{\ga}{\gamma}

\newcommand{\bra}[1]{\big< #1 \big|}
\newcommand{\ket}[1]{\big| #1 \big>}

\def \beq{\begin{equation}}
\def \eeq{\end{equation}}
\def \beqar{\begin{eqnarray}}
\def \eeqar{\end{eqnarray}}

\begin{document}
\bigskip
\bigskip
 \rightline{TUM/T39-97-6}
 \rightline{UFTP 439/1996}
\begin{center}
  {\bf Perturbative Part of the Non-Singlet Structure Function $F_2$ 
       in the Large-$N_F$ Limit}\footnote{Work
    supported in part by BMBF}\\
\end{center}
\bigskip
\centerline{\it L. Mankiewicz\footnote{On leave of absence from N. Copernicus
  Astronomical Center, Polish Academy of Science, ul. Bartycka 18, PL--00-716
  Warsaw (Poland)}} 
\centerline{\it Institute for Theoretical Physics} 
\centerline{\it TU-M\"unchen, D-85747 Garching, Germany} \medskip
 \medskip
\centerline{and}
\medskip
\centerline{\it  M.~Maul, E.~Stein}
\centerline{\it Institute for Theoretical Physics, 
Johann Wolfgang Goethe-Universit\"at,}
\centerline{\it D-69054 Frankfurt am Main, Germany} 

\begin{quote}
  We have calculated ${\overline{MS}}$ Wilson coefficients and anomalous
  dimensions for the non-singlet part of the structure function $F_2$ in the
  large-$N_F$ limit. Our result agrees with exact two and three loop
  calculations and gives the leading $N_F$ dependence of the perturbative
  non-singlet Wilson coefficients to all orders in $\alpha_S$.
\end{quote}
 \newpage 

\newpage High precision of current experimental data on deep-inelastic
lepton-nucleon scattering makes it necessary to extend perturbative
calculations to higher and higher orders.  Although in principle QCD
perturbation theory is well understood, in practice its technical complexity
has limited the existing state-of-the-art next to next to leading order (NNL)
calculations only to low Mellin moments $M_N, \, N=2,4,6,8$ of the structure
functions $F_2$ and $F_L$ \cite{LRV94,Lar96}.

Recently, it has been realized that in the large-$N_F$ limit of QCD
corresponding calculations can be performed exactly to all orders
\cite{Bro93,BB94}.  Although it is perhaps not easy to argue about the
importance of results obtained in this limit for the real world, they are
certainly useful either as an independent check of complicated computer algebra
algorithms used in the exact calculations, or as a starting point for further
approximation procedures, such as `Naive Nonabelianization' (NNA) \cite{NNA} or
`Asymptotic Pade Approximation' \cite{APA}.

The goal of the present paper is to analyze moments ${\cal
M}_{2,N}(Q^2)$ of the flavor non-singlet (NS) component of twist-2
part of structure function $F_2(x,Q^2)$
\begin{equation}
{\cal M}_{2,N}(Q^2) = \int_0^1 dx \, x^{N-2} F_2^{NS}(x,Q^2) \quad N =
2,4,\dots \, ,
\end{equation}
defined through the well known decomposition \cite{QCDbook} of the hadronic
scattering tensor of unpolarized deep inelastic lepton nucleon scattering in
terms of the structure functions $F_2(x,Q^2)$ and $F_L(x,Q^2)$.
\beqar
W_{\mu\nu}(p,q) &=& \frac{1}{2\pi}\,\int d^4z e^{i q z } 
\bra{p} J_\mu(z) J_\nu(0) \ket{p}
= \left(\g_{\mu\nu} - \frac{q_\mu q_\nu}{q^2} \right) 
\frac{1}{2x} F_L(x,Q^2) \nn \\
& - &
\left(\g_{\mu\nu} + p_\mu p_\nu \frac{q^2}{(p\cdot q)^2} 
- \frac{p_\mu q_\nu + p_\nu q_\mu}{p \cdot q} \right)\frac{1}{2 x}
F_2(x,Q^2) \, .
\nn \\
\label{Wmunu}
\eeqar
According to the Operator Product Expansion \cite{Yndbook} one
can arrange ${\cal M}_{2,N}(Q^2)$ as a product
\begin{equation}
{\cal M}_{2,N}(Q^2) = C_{2,N}(\alpha_S(Q^2)) A_N(Q^2)
\label{ope1}
\end{equation}
of the Wilson coefficient $C_{2,N}(\alpha_S(Q^2))$ and the
spin-averaged matrix element $A_N(Q^2)$ of a spin-N, twist-2 operator
\beq 
\bra{p} \bar{\psi} \ga^{\{\mu_1} i
D^{\mu_2} \ldots i D^{\mu_N\}} \psi \ket{p}^{p-n} = p^{\{\mu_1 \ldots}
p^{\mu_N\}} A_N(\mu^2) \quad.
\label{ope2}
\eeq
In the above equation {\small $\{\mu \dots \nu\}$} indicates twist-2
projection i.e., symmetric and traceless combination.  Validity of
equation (\ref{ope1}) requires that both renormalization and
factorization scales have been set equal to the virtuality of the
external photon $Q^2$.  In addition, it has been implicitly
understood that squares of quark charges are properly included on the
right-hand side of Eq.(\ref{ope2}), and that flavors of
quark-operators $\psi$ have been combined to yield the difference
between u- and d-quark matrix elements.

The $Q^2$-dependence of matrix elements $A_N(Q^2)$, written as a
solution of the renormalisation group equation,
\begin{equation}
A_N(Q^2) = A_N(\mu^2) \cdot 
\exp \left( \int_{\alpha_S(\mu^2)}^{\alpha_S(Q^2)} d {\alpha_S}'
\frac{\gamma_N^{(NS)}({\alpha_S}') }{\beta ({\alpha_S}')} \right) \quad.
\label{reno}
\end{equation}
is determined by a set of anomalous dimensions $\gamma_N(\alpha_S(Q^2))$.
The perturbative expansion of $C_{2,N}$ and $\gamma_N$ results in
\begin{eqnarray}
C_{2,N}(\alpha_S(Q^2)) & = & 1 + \sum_{j=0}^\infty C_{2,N}^{(j)}
a_s^{j+1} \quad,
\nonumber \\
\gamma_N(\alpha_S(Q^2)) & = & \sum_{j=0}^\infty \gamma_{N}^{(j)} a_s^{j+1}
\quad,
\label{pert}
\end{eqnarray}
where we have introduced a shorthand notation $a_s = \frac{\alpha_S(Q^2)}{4
  \pi}$.  In the following we present results for the Wilson coefficients
$C_{2,N }(\alpha_S(Q^2))$ and anomalous dimensions $\gamma_N(\alpha_S(Q^2))$
obtained in the large-$N_F$ limit, $N_F \to \infty$, keeping $a_s \cdot N_F =
{\rm const}$.

\begin{figure}
\psfig{figure=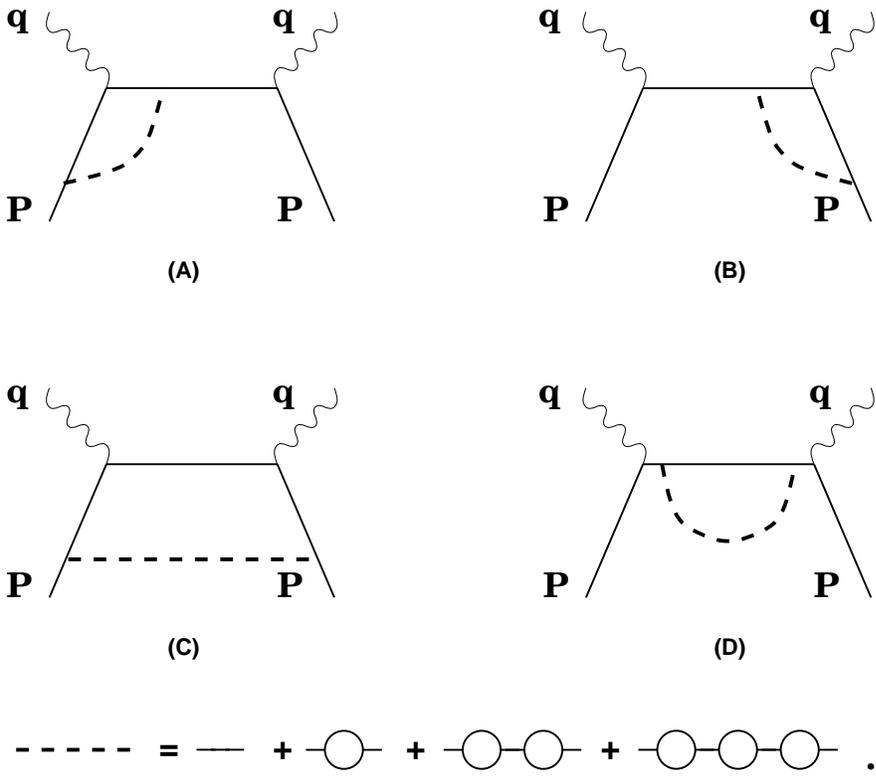,height=4in}
\caption{Graphs $A,B,C$ and $D$ which contribute to the calculation of
the perturbative part of $F_2(x,Q^2)$ in the large-$N_F$ limit.}
\label{fig1}   
\end{figure}

Technically, we have found it advantageous to follow the approach developed in
\cite{BB94,BBB95} and calculate the generating function $G(u;N)$
\begin{equation}
C_{2,N}^{(j)} =  \left(-\frac{2}{3} N_F\right)^j\left.\frac{d^j}{d u^j} 
                 \right|_{u=0} G(u;N) + {\cal O}(N_F^{j-1}) 
\, ,
\label{defG1}
\end{equation}
for the Wilson coefficients $C_{2,N}$ in the large-$N_F$ limit. As explained in
details e.g. in \cite{Bro93,BB94}, the dominant contribution in this limit
arises from an arbitrary number of fermionic loop insertions into the gluon
propagator.  As a consequence, the problem can be reduced to calculation of the
first-order radiative corrections using modified Landau-gauge gluon propagator
\begin{equation}
D_{\mu\nu}(k;u) = \frac{i}{(-k^2)^{1+u}} 
\left( g_{\mu\nu} - \frac{k_\mu k_\nu}{k^2} \right) \, .
\label{prop}
\end{equation}
In calculations of gauge invariant quantities, such as in the present
case, the longitudinal part does not contribute and can be
dropped. The only complication arises due to renormalization - the
`bare' function $G_B(u;N)$ acquires a singularity at $u = 0$ and has
to be renormalized by an appropriate counterterm. As discussed in
\cite{BB94,BBB95}, in minimal subtraction schemes such as
${\overline{MS}}$ the counterterm can be obtained in a compact form.
As a result the renormalized generating function $G_R(u;N)$ can be
written as
\begin{equation}
G_R(u;N) = G_B(u;N) + {\tilde G}_0(u;N)/u \quad,
\end{equation} 
where ${\tilde G}_0(u;N)$ is determined in terms of expansion
coefficients $g_k(N)$ of another function $G_0(u;N)$ in the variable
$u$ around the origin
\begin{eqnarray}
G_0(u;N) & = & \sum_{k=0}^\infty g_k(N) u^k \quad, \nonumber \\
{\tilde G}_0(u;N) & = & \sum_{k=0}^\infty \frac{1}{k!} g_k(N) u^k \, .
\label{Gtilde}
\end{eqnarray}
The merit of this complicated and apparently indirect construction lies in the
fact that the function $G_0(u;N)$ can be found directly from the calculation of
one-loop radiative corrections computed in d-dimensions using the modified
propagator (\ref{prop}), while the anomalous dimension follows from the 
corresponding counterterm in a standard way. An interested reader should
consult Refs. \cite{BB94,BBB95}, in particular Appendix A of Ref. \cite{BB94},
for details. To compute $G_R(u;N)$ and $\gamma_N$ we have calculated 
the Feynman diagrams depicted on Figure \ref{fig1}, arriving at
\begin{eqnarray}
G_R(u;N) & = & G_R^A(u;N) + G_R^B(u;N) + G_R^C(u;N) + G_R^D(u;N) \nonumber \\
G_R^A(u;N) & = & G_R^B(u;N) = 
2 C_F[\exp(-C)]^u   \frac{\Gamma(1-u)}{
               u \Gamma(3-u) (1+u)}
\nonumber \\
&&   + \sum_{k=2}^{N} 2C_F (\exp[-C])^u   
\frac{\Gamma(1-u)}{\Gamma(1+u) \Gamma(3-u)}
          \frac{\Gamma(u+k-1)}{\Gamma(k) (u+k)}
\nonumber\\
&&-2 C_F[\exp(-C)]^u \frac{ \Gamma(1-u) \Gamma(u+N)}{
               \Gamma(3-u) \Gamma(1+u) \Gamma(N) (u+N) }
\nonumber \\               
&&-\sum_{k=1}^N 2 C_F (\exp[-C])^u \frac{\Gamma(1-u)}{\Gamma(3-u) \Gamma(1+u)}
                 \frac{\Gamma(u+k)}{\Gamma(k) (u+k)}
\nonumber \\
&& +\sum_{k=1}^{N-1}  4 C_F (\exp[-C])^u \frac{\Gamma(1-u)}
            {u \Gamma(3-u) \Gamma(1+u)}
        \frac{\Gamma(u+k)}{\Gamma(k)(u+k+1)} \nonumber \\
&&+ \frac{{\tilde G}_0^A(u;N)}{u}
\nonumber\\
G_0^A(u;N) & = &-  \sum_{k=2}^{N} \frac{C_F}{3} \frac{ \Gamma(4+2u) }
                    { \Gamma(1-u)\Gamma^2(2+u) \Gamma(1+u)(u+k) }
\nonumber\\
&& -\frac{C_F}{3} \frac{ \Gamma(4+2u) }
                          { \Gamma(1-u)\Gamma^2(1+u) \Gamma(3+u) }  
\nonumber \\
\nonumber \\
G_R^C(u;N) & = & 
12 C_F[\exp(-C)]^u \frac{\Gamma(1-u) \Gamma(N+u)}
             { \Gamma(3-u)( 1+u+N) \Gamma(N) \Gamma(1+u)}
\nonumber \\
&&-2C_F [\exp(-C)]^u  
\frac{ \Gamma(1-u) (u^2 - 3u + 2) \Gamma^2(N+u)}
       {u \Gamma(3-u) \Gamma(1+u) \Gamma(N+u+2) \Gamma(N)} 
\nonumber \\
&&+ \frac{{\tilde G}_0^C(u;N)}{u}
\nonumber \\
G_0^C(u;N) & = &
\frac{C_F}{3} \frac{ \Gamma(4+2u)\Gamma(N+u) }
                          { \Gamma(1-u)\Gamma^3(1+u) \Gamma(N+u+2) } 
\nonumber \\
\nonumber \\
G_R^D(u;N) & = & 
-2C_F [\exp(-C)]^u 
\frac{  \Gamma(1-u) \Gamma(N+u)}{u\Gamma(3-u) \Gamma(1+u) \Gamma(N)} 
\nonumber \\
&&+ \frac{{\tilde G}_0^D(u;N)}{u}
\nonumber \\
G_0^D(u;N) & = & \frac{C_F}{3} \frac{ \Gamma(4+2u) }
                          { \Gamma(1-u)\Gamma^2(1+u) \Gamma(3+u) } 
\label{resG}
\end{eqnarray}
for the generating function $G_R(u;n)$. In the above formula $C_F = 4/3$, and
$C$ denotes the finite part of the quark loop insertion into the gluon
propagator, $C_{\overline {MS}} = - 5/3$. The relation between ${\tilde
  G}_0(u;N)$ and $G_0(u;N)$ is given by equation (\ref{Gtilde}). For the case
of the anomalous dimension $\gamma_{N}(a_S)$ our result reads
\begin{eqnarray}
\gamma_{N}(a_S) & = & \gamma_{N}^A(a_S) + \gamma_{N}^B(a_S) + \gamma_{N}^C(a_S)
 +
\gamma_{N}^D(a_S) \nonumber \\
\gamma_{N}^A(a_S) & = & \gamma_{N}^B(a_S) = - a_S C_F \frac{1}{3} 
\frac{ \Gamma(4+2s) }{ \Gamma(1-s)\Gamma^2(1+s) \Gamma(3+s) } 
\nonumber \\
&&
-a_S C_F  \sum_{k=2}^{N} \frac{1}{3} \frac{ \Gamma(4+2s) }
                    { \Gamma(1-s)\Gamma^2(2+s) \Gamma(1+s)(s+k) } 
 \nonumber \\
\gamma_{N}^C(a_S) & = & a_S C_F  
        \frac{1}{3} \frac{ \Gamma(4+2s)\Gamma(N+s) }
                          { \Gamma(1-s)\Gamma^3(1+s) \Gamma(N+s+2) } 
 \nonumber \\
\gamma_{N}^D(a_S) & = & a_S C_F  
\frac{1}{3} \frac{ \Gamma(4+2s) }
                          { \Gamma(1-s)\Gamma^2(1+s) \Gamma(3+s) } \, ,
\nonumber \\
\label{resGamma}
\end{eqnarray}
where $s = \left(- \frac{2}{3} N_F \right) a_s$. Terms labeled as $A,B,C$ and
$D$ in Eqs. (\ref{resG}) and (\ref{resGamma}) correspond to contributions of
graphs depicted in Figure \ref{fig1}. We quote results obtained in the Feynman
gauge i.e. contributions from the longitudinal part of the propagator
(\ref{prop}), which cancel in the sum of all graphs, have been neglected. In
the case of the anomalous dimension $\gamma_N(a_s)$ one can put together
various terms in (\ref{resGamma}) and obtain a more compact expression
\begin{eqnarray}
\gamma_N(a_s) &=& a_s C_F \frac{1}{3} 
\frac{\Gamma(4+2s)}{\Gamma(1-s)\Gamma(1+s)^3}
\nonumber \\
\nonumber \\
&&\times \left[ \frac{1}{(s+2)(s+1)}
       - \frac{1}{(s+N+1)(s+N)} 
      + \frac{2}{(1+s)^2}\sum_{k=2}^N \frac{1}{k+s} 
\right] \, ,
\nonumber \\
\label{gracey}
\end{eqnarray} 
$s = \left(- \frac{2}{3} N_F \right) a_s$, which agrees with result
obtained earlier by Gracey \cite{Gra95}.
\newline
Now, expanding the right-hand side of (\ref{gracey}) in $a_s$ 
\begin{equation}
\gamma_N(a_s) = b_1 C_F a_s + C_F \sum_{j=2}
\left[b_j\left(-\frac{2}{3} N_F\right)^{j-1}a_s^j
\right]
\end{equation}
one can find explicit results for the anomalous dimensions in any fixed order
of the perturbative expansion in the large-$N_F$-limit, see Ref. \cite{Gra95}.
Similarly, using (\ref{defG1}) one can derive from Eq.~(\ref{resG}) fixed order
predictions for the Wilson coefficients for the structure function $F_2$ in
this limit. Note that large-$N_F$ generating function for the Wilson
coefficients for the non-singlet part of the structure function $F_L(x,Q^2)$ is
known from Refs. \cite{Gra95,SMMS96}.

In Ref. \cite{Lar96} Wilson coefficients for $F_L$ and $F_2$ have been
calculated for the even moments $N = 2 \dots 8$ exactly up to the third order
in $\alpha_s$.  The leading $N_F$ terms at the third and the fourth orders for
an arbitrary $N$-th moment read
\beqar
C_{2,N}^{(3)} &=& 
C_F \left(-\frac{2}{3}N_F\right)^2\bigg[
 - {\displaystyle \frac {9517}{216}}  - 2
\,\zeta (3) + {\displaystyle \frac {1}{81}} \,{\displaystyle 
\frac {5659 + 108\,\zeta (3)}{1 + N}}  - {\displaystyle \frac {
368}{9}} \,{\displaystyle \frac {1}{(1 + N)^{2}}}  + 
{\displaystyle \frac {146}{9}} \,{\displaystyle \frac {1}{(1 + N)
^{3}}} 
\nn \\
 && - {\displaystyle \frac {10}{3}} \,{\displaystyle \frac {1
}{(1 + N)^{4}}} 
  + {\displaystyle \frac {1}{81}} \,{\displaystyle 
\frac { - 2041 + 108\,\zeta (3)}{N}}  - {\displaystyle \frac {30
}{N^{2}}}  + {\displaystyle \frac {134}{9}} \,{\displaystyle 
\frac {1}{N^{3}}}  - {\displaystyle \frac {10}{3}} \,
{\displaystyle \frac {1}{N^{4}}}  + {\displaystyle \frac {4955}{
81}} \,{\displaystyle \frac {1}{N}} 
\nn  \\
 &&  + ({\displaystyle \frac {8}{3}} \,\zeta (3) + 
{\displaystyle \frac {4357}{162}}  - {\displaystyle \frac {4}{N^{
3}}}  + {\displaystyle \frac {32}{3}} \,{\displaystyle \frac {1}{
N^{2}}}  - {\displaystyle \frac {164}{9}} \,{\displaystyle 
\frac {1}{N}}  + {\displaystyle \frac {4}{(1 + N)^{3}}}  - 
{\displaystyle \frac {56}{3}} \,{\displaystyle \frac {1}{(1 + N)
^{2}}}  
\nn \\ & & 
+ {\displaystyle \frac {392}{9}} \,{\displaystyle \frac {
1}{1 + N}} )\,{\rm S_1}(N - 1) 
+ (1 + {\displaystyle \frac {2}{N}}  - 
{\displaystyle \frac {2}{1 + N}} )\,{\rm S_1}(N - 1)\,{\rm S_2}
(N - 1) 
\nn \\
 & &  + ( - {\displaystyle \frac {19}{6}}  + 
{\displaystyle \frac {2}{N^{2}}}  - {\displaystyle \frac {16}{3}
} \,{\displaystyle \frac {1}{N}}  - {\displaystyle \frac {2}{(1
 + N)^{2}}}  + {\displaystyle \frac {28}{3}} \,{\displaystyle 
\frac {1}{1 + N}} )\,{\rm S_1}(N - 1)^{2} 
\nn \\
 & &  + ( - {\displaystyle \frac {1}{3}}  - 
{\displaystyle \frac {2}{3}} \,{\displaystyle \frac {1}{N}}  + 
{\displaystyle \frac {2}{3}} \,{\displaystyle \frac {1}{1 + N}} )
\,{\rm S_1}(N - 1)^{3} 
\nn \\
 & &  + ( - {\displaystyle \frac {535}{18}}  - 
{\displaystyle \frac {2}{N^{2}}}  + {\displaystyle \frac {16}{3}
} \,{\displaystyle \frac {1}{N}}  + {\displaystyle \frac {2}{(1
 + N)^{2}}}  - {\displaystyle \frac {28}{3}} \,{\displaystyle 
\frac {1}{1 + N}} )\,{\rm S_2}(N - 1) 
\nn \\
 & &  + ({\displaystyle \frac {166}{9}}  - {\displaystyle 
\frac {4}{3}} \,{\displaystyle \frac {1}{N}}  + {\displaystyle 
\frac {4}{3}} \,{\displaystyle \frac {1}{1 + N}} )\,{\rm S_3}(N
 - 1) - {\displaystyle \frac {20}{3}} \,{\rm S_4}(N - 1) 
\nn \\
 & &  +    \! {\displaystyle \sum _{k=1}^{N - 1}} \,
\left( - {\displaystyle \frac {2}{9}} \,{\displaystyle \frac {( - 146
\,k^{4} - 139\,k^{3} - 8\,k^{2} + 3\,k - 18)\,{\rm S_1}(k)}{(k
 + 1)^{3}\,k^{2}}}  \!  \right)  
\nn \\
 & &  +  \! {\displaystyle \sum _{k=1}^{N - 1}} \,
\left( - {\displaystyle \frac {2}{9}} \,{\displaystyle \frac {(18\,k^{
4} + 36\,k^{3} + 18\,k^{2})\,{\rm S_1}(k)\,{\rm S_2}(k)}{(k + 1
)^{3}\,k^{2}}}  \!  \right)  
\nn \\
 & &  +   \! {\displaystyle \sum _{k=1}^{N - 1}} \,
{\displaystyle \frac {2}{3}} \,{\displaystyle \frac {(16\,k^{2}
 + 7\,k - 3)\,{\rm S_1}(k)^{2}}{k\,(k + 1)^{2}}}  \!   
 +  \! {\displaystyle \sum _{k=1}^{N - 1}} 
{\displaystyle \frac {4}{3}} \,{\displaystyle \frac {{\rm S_1}(k
)^{3}}{k + 1}} 
\nn \\
 & &  -  \! {\displaystyle \sum _{k=1}^{N - 1}} \,
  {\displaystyle \frac {2}{3}} \,{\displaystyle \frac {(16\,k^{
2} + 7\,k - 3)\,{\rm S_2}(k)}{k\,(k + 1)^{2}}} +
{\displaystyle \sum _{k=1}^{N - 1}} \,
{\displaystyle \frac {8}{3}} \,{\displaystyle \frac {{\rm S_3}(k
)}{k + 1}} \bigg] +  {\cal O}(N_F)\;,
\nonumber \\
\nonumber \\
C_{2,N}^{(4)} &=& 
C_F \left(-\frac{2}{3}N_F\right)^3\Bigg[
{\displaystyle \frac {386}{3}} \,
{\displaystyle \frac {1}{(1 + N)^{3}}}  - {\displaystyle \frac {
325}{6}} \,{\displaystyle \frac {1}{(1 + N)^{4}}}  +  \left(  \! 
{\displaystyle \sum _{k=1}^{N - 1}} \,( - 6\,{\displaystyle 
\frac {{\rm S_4}(k)}{k + 1}} ) \!  \right)  - {\displaystyle 
\frac {1}{4}} \,\zeta (3) 
\nn \\ & & 
- {\displaystyle \frac {1}{27}} \,
{\displaystyle \frac {27\,\zeta (3) - 2068}{N^2}}  
 - {\displaystyle \frac {1}{27}} \,{\displaystyle 
\frac {5713 + 27\,\zeta (3)}{(1 + N)^{2}}}  - {\displaystyle 
\frac {4955}{27}} \,{\displaystyle \frac {1}{N^{2}}}  + 
{\displaystyle \frac {83}{N^{3}}}  - {\displaystyle \frac {265}{6
}} \,{\displaystyle \frac {1}{N^{4}}}  + {\displaystyle \frac {
34883}{162}} \,{\displaystyle \frac {1}{N}}  
\nn \\
 & &   + {\displaystyle \frac {1}{1620}} \,{\displaystyle 
\frac {452200 + 5940\,\zeta (3) - 2430\,\zeta (4)}{1 + N}}  - 
{\displaystyle \frac {1}{1620}} \,{\displaystyle \frac {540\,
\zeta (3) + 154360 + 2430\,\zeta (4)}{N}}  + {\displaystyle 
\frac {9}{4}} \,\zeta (4) 
\nn \\
 & &   + {\displaystyle \frac {23}{2}} \,{\displaystyle 
\frac {1}{N^{5}}}  + {\displaystyle \frac {23}{2}} \,
{\displaystyle \frac {1}{(1 + N)^{5}}}  + ( - 3\,\zeta (4) + 
{\displaystyle \frac {10}{3}} \,\zeta (3) + {\displaystyle 
\frac {25279}{324}}  + {\displaystyle \frac {12}{N^{4}}}  - 
{\displaystyle \frac {32}{N^{3}}}  + {\displaystyle \frac {164}{3
}} \,{\displaystyle \frac {1}{N^{2}}}  
\nn \\ & &   
- {\displaystyle \frac {
2077}{27}} \,{\displaystyle \frac {1}{N}}  
- {\displaystyle \frac {12}{(1 + N)^{4}}}  + 
{\displaystyle \frac {56}{(1 + N)^{3}}}  - {\displaystyle \frac {
392}{3}} \,{\displaystyle \frac {1}{(1 + N)^{2}}}  + 
{\displaystyle \frac {5731}{27}} \,{\displaystyle \frac {1}{1 + N
}} ){\rm S_1}(N - 1) 
\nn \\
 & &   + ({\displaystyle \frac {19}{2}}  - {\displaystyle 
\frac {6}{N^{2}}}  + {\displaystyle \frac {16}{N}}  + 
{\displaystyle \frac {6}{(1 + N)^{2}}}  - {\displaystyle \frac {
28}{1 + N}} )\,{\rm S_1}(N - 1)\,{\rm S_2}(N - 1) 
\nn \\
 & &   + ( - {\displaystyle \frac {1}{4}}  - 
{\displaystyle \frac {1}{2}} \,{\displaystyle \frac {1}{N}}  + 
{\displaystyle \frac {1}{2}} \,{\displaystyle \frac {1}{1 + N}} )
\,{\rm S_1}(N - 1)^{4} + ( - {\displaystyle \frac {3}{4}}  - 
{\displaystyle \frac {3}{2}} \,{\displaystyle \frac {1}{N}}  + 
{\displaystyle \frac {3}{2}} \,{\displaystyle \frac {1}{1 + N}} )
\,{\rm S_2}(N - 1)^{2} 
\nn \\
 & &   + ( - {\displaystyle \frac {203}{12}}  - 
{\displaystyle \frac {6}{N^{3}}}  + {\displaystyle \frac {16}{N^{
2}}}  - {\displaystyle \frac {82}{3}} \,{\displaystyle \frac {1}{
N}}  + {\displaystyle \frac {6}{(1 + N)^{3}}}  - {\displaystyle 
\frac {28}{(1 + N)^{2}}}  + {\displaystyle \frac {196}{3}} \,
{\displaystyle \frac {1}{1 + N}} )\,{\rm S_1}(N - 1)^{2} 
\nn \\
 & &   + ( - {\displaystyle \frac {19}{6}}  + 
{\displaystyle \frac {2}{N^{2}}}  - {\displaystyle \frac {16}{3}
} \,{\displaystyle \frac {1}{N}}  - {\displaystyle \frac {2}{(1
 + N)^{2}}}  + {\displaystyle \frac {28}{3}} \,{\displaystyle 
\frac {1}{1 + N}} )\,{\rm S_1}(N - 1)^{3} 
\nn \\
 & &   + ({\displaystyle \frac {3}{2}}  + {\displaystyle 
\frac {3}{N}}  - {\displaystyle \frac {3}{1 + N}} )\,{\rm S_1}(N
 - 1)^{2}\,{\rm S_2}(N - 1) + ( - 2 - {\displaystyle \frac {4}{N
}}  + {\displaystyle \frac {4}{1 + N}} )\,{\rm S_1}(N - 1)\,
{\rm S_3}(N - 1) 
\nn \\
 & &   + ( - {\displaystyle \frac {10649}{108}}  - 2\,\zeta
 (3) + {\displaystyle \frac {6}{N^{3}}}  - {\displaystyle \frac {
16}{N^{2}}}  + {\displaystyle \frac {82}{3}} \,{\displaystyle 
\frac {1}{N}}  - {\displaystyle \frac {6}{(1 + N)^{3}}}  + 
{\displaystyle \frac {28}{(1 + N)^{2}}}  
\nn \\ & & 
- {\displaystyle \frac {
196}{3}} \,{\displaystyle \frac {1}{1 + N}} )\,{\rm S_2}(N - 1) 
+ ({\displaystyle \frac {274}{3}}  + {\displaystyle 
\frac {4}{N^{2}}}  - {\displaystyle \frac {32}{3}} \,
{\displaystyle \frac {1}{N}}  - {\displaystyle \frac {4}{(1 + N)
^{2}}}  + {\displaystyle \frac {56}{3}} \,{\displaystyle \frac {1
}{1 + N}} )\,{\rm S_3}(N - 1) 
\nn \\
 & &   +  {\displaystyle \sum _{k=1}^{N - 1}} \,
\left( - {\displaystyle \frac {1}{3}} \,{\displaystyle \frac {(146\,k
^{4} + 139\,k^{3} + 8\,k^{2} - 3\,k + 18)\,{\rm S_2}(k)}{k^{2}\,
(k + 1)^{3}}} \right)   +  {\displaystyle \sum _{
k=1}^{N - 1}} \,\left(3\,{\displaystyle \frac {{\rm S_2}(k)^{2}}{k + 
1}} \right)  
\nn \\
 & &   +  {\displaystyle \sum _{k=1}^{N - 1}} \,
\left({\displaystyle \frac {1}{3}} \,{\displaystyle \frac {(146\,k^{4}
 + 139\,k^{3} + 8\,k^{2} - 3\,k + 18)\,{\rm S_1}(k)^{2}}{k^{2}\,
(k + 1)^{3}}} \right) 
\nn \\
 & &   +  {\displaystyle \sum _{k=1}^{N - 1}} \,
\left( - {\displaystyle \frac {1}{27}} \,{\displaystyle \frac {(324 - 
3128\,k^{6} - 4929\,k^{5} - 2427\,k^{4} - 257\,k^{3} - 333\,k^{2}
 + 270\,k)\,{\rm S_1}(k)}{(k + 1)^{4}\,k^{3}}} \right)
\nn \\
 & &   +  {\displaystyle \sum _{k=1}^{N - 1}} \,
\left({\displaystyle \frac {2}{3}} \,{\displaystyle \frac {(16\,k^{2}
 + 7\,k - 3)\,{\rm S_1}(k)^{3}}{k\,(k + 1)^{2}}} \right) 
 +  {\displaystyle \sum _{k=1}^{N - 1}} \,
{\displaystyle \frac {{\rm S_1}(k)^{4}}{k + 1}}  - 
{\displaystyle \frac {281971}{1728}}  
\nn \\
 & &   + ( - {\displaystyle \frac {365}{6}}  + 
{\displaystyle \frac {3}{N}}  - {\displaystyle \frac {3}{1 + N}} 
)\,{\rm S_4}(N - 1) + 23\,{\rm S_5}(N - 1) 
\nn \\
 & &   +  {\displaystyle \sum _{k=1}^{N - 1}} \,
\left( - {\displaystyle \frac {1}{27}} \,{\displaystyle \frac {( - 216
\,k^{6} - 648\,k^{5} - 648\,k^{4} - 216\,k^{3})\,{\rm S_1}(k)\,
{\rm S_3}(k)}{(k + 1)^{4}\,k^{3}}} \right) 
\nn \\
 & &   +  {\displaystyle \sum _{k=1}^{N - 1}} \,
\left({\displaystyle \frac {4}{3}} \,{\displaystyle \frac {(16\,k^{2}
 + 7\,k - 3)\,{\rm S_3}(k)}{k\,(k + 1)^{2}}} \right) 
\nn \\
 & &   + {\displaystyle \sum _{k=1}^{N - 1}} \,
\left( - {\displaystyle \frac {1}{3}} \,{\displaystyle \frac {(18\,k^{
4} + 36\,k^{3} + 18\,k^{2})\,{\rm S_1}(k)^{2}\,{\rm S_2}(k)}{(k
 + 1)^{3}\,k^{2}}} \right) 
\nn \\
 & &   +  {\displaystyle \sum _{k=1}^{N - 1}} \,
\left( - {\displaystyle \frac {1}{27}} \,{\displaystyle \frac {(864\,k
^{6} + 2106\,k^{5} + 1458\,k^{4} + 54\,k^{3} - 162\,k^{2})\,{\rm 
S_1}(k)\,{\rm S_2}(k)}{(k + 1)^{4}\,k^{3}}} \right) 
\Bigg] + {\cal O}(N_F^2) \, , \nonumber \\
\eeqar
where $S_\alpha(N)= \sum_{k = 1}^N(1/k^\alpha)$.
The corresponding results for the longitudinal structure function
$F_L$  read
\beqar
C_{L,N}^{(3)} &=& \left(-\frac{2}{3}N_F\right)^2 \frac{4 C_F}{1+N}
\bigg[\frac{203}{18} - \frac{19}{3 (1+N)} + \frac{2}{(1+N)^2} +
\left(\frac{19}{3} - \frac{2}{(1+N)}\right)S_1(N-1) 
\nn \\
&&  + S_1^2(N-1)
- S_2(N-1)\bigg] + {\cal O}(N_F)\;,
\nn\\
C_{L,N}^{(4)} &=& \left(-\frac{2}{3}N_F\right)^3 \frac{C_F}{1+N}
\bigg[\frac{4955}{27} - \frac{406}{3 (1+N)} + \frac{76}{(1+N)^2} 
- \frac{24}{(1+N)^3} 
\nn \\
&&+ \left(\frac{406}{3} - \frac{76}{(1+N)} + \frac{24}{(1+N)^2}\right)S_1(N-1) 
+
\left(38 - \frac{12}{1+N}\right) S_1^2(N-1) 
\nn \\
&&
+ 4 S_1^3(N-1)
 - \left(38 - \frac{12}{(1+N)}+ 12 S_1(N-1)\right) S_2(N-1) 
\nn \\
&&
+ 8 S_3(N-1)\bigg] + {\cal O}(N_F^2)\;,
\nn \\
C_{L,N}^{(5)}  &=& \left(-\frac{2}{3}N_F\right)^4 \frac{C_F}{1+N}
\Bigg[\frac{69766}{81} - \frac{19820}{27(1+N)} + \frac{1624}{3(1+N)^2}
- \frac{304}{(1+N)^3} + 96 \frac{1}{(1+N)^4}
\nn \\
&& 
+ \bigg(\frac{19820}{27} - \frac{1624}{3(1+N)} + \frac{304}{(1+N)^2}
- 96 \frac{1}{(1+N)^3} \nn \\
&& - (152 - \frac{48}{(1+N)}) S_2(N-1) + 32 S_3(N-1) \bigg)
S_1(N-1)\nn \\
&&+ 
\bigg(\frac{812}{3} - \frac{152}{1+N} + \frac{48}{(1+N)^2} -
24 S_2(N-1)\bigg) S_1^2(N-1)
\nn \\
&&
+ \bigg(\frac{152}{3} - \frac{16}{1+N}\bigg) S_1^3(N-1) + 4 S_1^4(N-1)
\nn \\
&& 
- \bigg(\frac{812}{3} - \frac{152}{(1+N)} +
\frac{48}{(1+N)^2}\bigg)S_2(N-1)
+ 12 S_2^2(N-1)
\nn \\
&& + \bigg(\frac{304}{3} - \frac{32}{1+N}\bigg)S_3(N-1) - 24 S_4(N-1)
\Bigg] + {\cal O}(N_F^3) \, ,
\eeqar
respectively. We have checked that the first, second (not quoted here) and the
third order coefficients agree with the leading-$N_F$ terms extracted from Ref.
\cite{Lar96}.  Results for higher order terms are too long to be written down
explicitly.  Instead, we have prepared a Maple program for the numerical
evaluation of Wilson coefficients in the large-$N_F$ limit, available on 
request, which allows to compute them using results of \cite{Gra95,SMMS96} 
and of the present paper.

\begin{table}
\renewcommand{\arraystretch}{1.3}
\begin{center}
\begin{tabular}{l|l|l|l}
$N_F = 3$      &  Exact results \cite{Lar96} &   NNA approximants \\ \hline
$N = 2$ 
& $ 1.69377 \;a_s^2 + 1.42209 \;a_s^3$  
             & $71.99999 \;a_s^2  + 1099.02 \;a_s^3 + 26193.11796\;a_s^4$ \\
$N = 4$ 
& $ 91.3797 \;a_s^2 + 1675.76 \;a_s^3$
             & $229.3368 \;a_s^2 + 4256.46\;a_s^3 + 103655.6199\;a_s^4$ \\
$N = 6$ 
& $ 218.3596 \;a_s^2 +  5004.63 \;a_s^3$
             & $ 378.1762 \;a_s^2 + 7775.28 \;a_s^3 + 200938.7405 \;a_s^4$ \\
$N = 8$ 
& $ 357.0330 \;a_s^2 +  9357.69 \;a_s^3$
             & $ 511.9851 \;a_s^2 +   11283.4 \;a_s^3 + 306602.1048 \;a_s^4$\\
\end{tabular}
\end{center}
\caption[]{\sf
Comparison of the NNA approximants to the exact results of the coefficient
function $C_{2,N}(a_s)$ obtained in \cite{Lar96} up to order ${\cal O} (a_s^3)$
for $N_F = 3$. The last column contains also the NNA prediction for the ${\cal
O}(a_s^4)$ terms. The ${\cal O}(a_s)$ corrections agree of course exactly.}
\label{renf2N3}
\renewcommand{\arraystretch}{1.0}

\renewcommand{\arraystretch}{1.3}
\begin{center}
\begin{tabular}{l|l|l|l}
$N_F = 4$         &  Exact results \cite{Lar96} &   NNA approximants \\ \hline
$N = 2$ &
 $-3.63957 \;a_s^2 - 169.747. \;a_s^3$
             & $66.66666 \;a_s^2 + 942.230 \;a_s^3 + 20792.94153\;a_s^4$  \\ \
$N = 4$ &
 $ 74.3918 \;a_s^2 + 901.570 \;a_s^3$
            & $212.3489 \;a_s^2 + 3649.22  \;a_s^3 + 82285.17302\;a_s^4$ \\ 
$N = 6$ &
$ 190.3465 \;a_s^2 + 3454.66 \;a_s^3$
             & $ 350.1631 \;a_s^2  + 6666.05 \;a_s^3 +159511.6507 \;a_s^4$ \\ 
$N = 8$ &
 $ 319.1081 \;a_s^2 + 6973.59 \;a_s^3$
             & $ 474.0603 \;a_s^2   + 9673.72 \;a_s^3 +243390.6360 \;a_s^4$\\ 
\end{tabular}
\end{center}
\caption[]{\sf
Comparison of the NNA approximants to the exact results of the coefficient func
tion
$C_{2,N}(a_s)$ obtained in \cite{Lar96} up to order ${\cal O} (a_s^3)$ for $N_F
= 4$. The last column contains also the prediction for the ${\cal O}(a_s^4)$
terms.  The ${\cal O}(a_s)$ corrections agree of course exactly.} 
\label{renf2N4}
\renewcommand{\arraystretch}{1.0}
\end{table}

Finally, in Tables \ref{renf2N3} and \ref{renf2N4} we have compared the NNA
approximants for the coefficient functions $C_{2,N}(a_s)$ with the exact
results obtained in Ref. \cite{Lar96}, and evaluate the NNA prediction for the
${\cal O}(a_s^4)$ terms. It is seen that although typically the NNA procedure
predicts correctly the magnitude of the perturbative coefficients, with
increasing precision as $N$ becomes larger, the numerical accuracy is not
optimal. We have also checked that a similar procedure applied to the anomalous
dimension $\gamma_N(a_s)$ gives much worse results which is probably connected
to the fact that in the latter case the perturbative expansion is not
dominated by renormalons.

\vskip .7 cm

{\bf Acknowledgment:} We are grateful to V. Braun and M. Beneke for useful
discussions. This work has been supported by BMBF, DFG (G.~Hess Programm),
DESY, and German-Polish exchange program X081.91. L.~M.~was supported in part
by the KBN grant 2~P03B~065~10.

\newpage
\def \ajp#1#2#3{Am.~J.~Phys.~{\bf#1} (#3) #2}
\def \apny#1#2#3{Ann.~Phys.~(N.Y.) {\bf#1} (#3) #2}
\def \app#1#2#3{Acta Phys.~Polonica {\bf#1} (#3) #2 }
\def \arnps#1#2#3{Ann.~Rev.~Nucl.~Part.~Sci.~{\bf#1} (#3) #2}
\def \cmp#1#2#3{Commun.~Math.~Phys.~{\bf#1} (#3) #2}
\def \cmts#1#2#3{Comments on Nucl.~Part.~Phys.~{\bf#1} (#3) #2}
\def \cn{Collaboration}
\def \corn93{{\it Lepton and Photon Interactions:  XVI International Symposium,
Ithaca, NY August 1993}, AIP Conference Proceedings No.~302, ed.~by P. Drell
and D. Rubin (AIP, New York, 1994)}
\def \cp89{{\it CP Violation,} edited by C. Jarlskog (World Scientific,
Singapore, 1989)}
\def \dpff{{\it The Fermilab Meeting -- DPF 92} (7th Meeting of the American
Physical Society Division of Particles and Fields), 10--14 November 1992,
ed. by C. H. Albright \ite~(World Scientific, Singapore, 1993)}
\def \dpf94{DPF 94 Meeting, Albuquerque, NM, Aug.~2--6, 1994}
\def \efi{Enrico Fermi Institute Report No. EFI}
\def \el#1#2#3{Europhys.~Lett.~{\bf#1} (#3) #2}
\def \f79{{\it Proceedings of the 1979 International Symposium on Lepton and
Photon Interactions at High Energies,} Fermilab, August 23-29, 1979, ed.~by
T. B. W. Kirk and H. D. I. Abarbanel (Fermi National Accelerator Laboratory,
Batavia, IL, 1979}
\def \hb87{{\it Proceeding of the 1987 International Symposium on Lepton and
Photon Interactions at High Energies,} Hamburg, 1987, ed.~by W. Bartel
and R. R\"uckl (Nucl. Phys. B, Proc. Suppl., vol. 3) (North-Holland,
Amsterdam, 1988)}
\def \ib{{\it ibid.}~}
\def \ibj#1#2#3{~{\bf#1} (#3) #2}
\def \ichep72{{\it Proceedings of the XVI International Conference on High
Energy Physics}, Chicago and Batavia, Illinois, Sept. 6--13, 1972,
edited by J. D. Jackson, A. Roberts, and R. Donaldson (Fermilab, Batavia,
IL, 1972)}
\def \ijmpa#1#2#3{Int.~J.~Mod.~Phys.~A {\bf#1} (#3) #2}
\def \ite{{\it et al.}}
\def \jmp#1#2#3{J.~Math.~Phys.~{\bf#1} (#3) #2}
\def \jpg#1#2#3{J.~Phys.~G {\bf#1} (#3) #2}
\def \lkl87{{\it Selected Topics in Electroweak Interactions} (Proceedings of
the Second Lake Louise Institute on New Frontiers in Particle Physics, 15--21
February, 1987), edited by J. M. Cameron \ite~(World Scientific, Singapore,
1987)}
\def \ky85{{\it Proceedings of the International Symposium on Lepton and
Photon Interactions at High Energy,} Kyoto, Aug.~19-24, 1985, edited by M.
Konuma and K. Takahashi (Kyoto Univ., Kyoto, 1985)}
\def \mpla#1#2#3{Mod.~Phys.~Lett.~A {\bf#1} (#3) #2}
\def \nc#1#2#3{Nuovo Cim.~{\bf#1} (#3) #2}
\def \npb#1#2#3{Nucl.~Phys.~B {\bf#1} (#3) #2}
\def \pisma#1#2#3#4{Pis'ma Zh.~Eksp.~Teor.~Fiz.~{\bf#1} (#3) #2[JETP Lett.
{\bf#1} (#3) #4]}
\def \pl#1#2#3{Phys.~Lett.~{\bf#1} (#3) #2}
\def \plb#1#2#3{Phys.~Lett.~B {\bf#1} (#3) #2}
\def \pr#1#2#3{Phys.~Rev.~{\bf#1} (#3) #2}
\def \pra#1#2#3{Phys.~Rev.~A {\bf#1} (#3) #2}
\def \prd#1#2#3{Phys.~Rev.~D {\bf#1} (#3) #2}
\def \prl#1#2#3{Phys.~Rev.~Lett.~{\bf#1} (#3) #2}
\def \prp#1#2#3{Phys.~Rep.~{\bf#1} (#3) #2}
\def \ptp#1#2#3{Prog.~Theor.~Phys.~{\bf#1} (#3) #2}
\def \rmp#1#2#3{Rev.~Mod.~Phys.~{\bf#1} (#3) #2}
\def \rp#1{~~~~~\ldots\ldots{\rm rp~}{#1}~~~~~}
\def \si90{25th International Conference on High Energy Physics, Singapore,
Aug. 2-8, 1990}
\def \slc87{{\it Proceedings of the Salt Lake City Meeting} (Division of
Particles and Fields, American Physical Society, Salt Lake City, Utah, 1987),
ed.~by C. DeTar and J. S. Ball (World Scientific, Singapore, 1987)}
\def \slac89{{\it Proceedings of the XIVth International Symposium on
Lepton and Photon Interactions,} Stanford, California, 1989, edited by M.
Riordan (World Scientific, Singapore, 1990)}
\def \smass82{{\it Proceedings of the 1982 DPF Summer Study on Elementary
Particle Physics and Future Facilities}, Snowmass, Colorado, edited by R.
Donaldson, R. Gustafson, and F. Paige (World Scientific, Singapore, 1982)}
\def \smass90{{\it Research Directions for the Decade} (Proceedings of the
1990 Summer Study on High Energy Physics, June 25 -- July 13, Snowmass,
Colorado), edited by E. L. Berger (World Scientific, Singapore, 1992)}
\def \stone{{\it B Decays}, edited by S. Stone (World Scientific, Singapore,
1994)}
\def \tasi90{{\it Testing the Standard Model} (Proceedings of the 1990
Theoretical Advanced Study Institute in Elementary Particle Physics, Boulder,
Colorado, 3--27 June, 1990), edited by M. Cveti\v{c} and P. Langacker
(World Scientific, Singapore, 1991)}
\def \yaf#1#2#3#4{Yad.~Fiz.~{\bf#1} (#3) #2 [Sov.~J.~Nucl.~Phys.~{\bf #1} (#3)
#4]}
\def \zhetf#1#2#3#4#5#6{Zh.~Eksp.~Teor.~Fiz.~{\bf #1} (#3) #2 [Sov.~Phys. -
JETP {\bf #4} (#6) #5]}
\def \zpc#1#2#3{Zeit.~Phys.~C {\bf#1} (#3) #2}

\newpage

\end{document}